\documentclass[runningheads,a4paper]{llncs}
\usepackage{graphicx}
\usepackage{amssymb}

\begin{document}

\title{Geo-aggregation permits low stretch and routing tables of logarithmical size}
\author{Victor S. Grishchenko}
\institute{Ural State University \\ \email { gritzko@ural.ru }}

\maketitle

\begin{abstract}
This article first addresses applicability of Euclidean models to the domain of Internet routing. Those models are found (limitedly) applicable. Then a simplistic model of routing is constructed for Euclidean plane densely covered with points-routers. The model guarantees low stretch and logarithmical size of routing tables at any node. The paper concludes with a discussion on applicability of the model to real-world Internet routing.
\end{abstract}

\section {Applicability of the model }

The underlying paradigm of the studied model is routing on a Euclidean plane. Applicability of this model is not obvious. On the contrary. First, it is a common understanding that Internet eliminates distance. Second, Internet is usually modeled as a discrete graph, not a solid plane. I shall consider both aspects.

\subsection {Distance}

It is a known observation \cite{skitter} that round trip time ($\tau$) reliably correlates with geographical distance. RTT has an obvious lower bound, $\tau \ge \tau_{c} = \frac{2d}{c}$, where $d$ is  distance and $c$ is speed of light. Although the way RTT depends on distance is rather complicated, practical RTT usually has the same order of magnitude as its lower bound. (From my personal limited experience, usually $\tau_{c} < \tau < 10 \tau_{c}$ for destinations farther than 1,000km).

Table~\ref{tab:rtt} shows that RTTs for different paths across the globe may differ by three orders of magnitude. It is a too significant difference to be neglected.

I'll define a indistinguishability distance as $d_{i18y} = \frac{1}{2} c \tau_{0}$, where $\tau_{o}$ is ``zero RTT'' (``ping localhost RTT''). Currently, $d_{i18y} \approx 0.5\cdot3.0\cdot10^{5}\cdot10^{-3} = 150km$. Due to the difference between $\tau$ and $\tau_{c}$ we may further lower this estimation to ``practical i18y distance'' of $\sim50km$. This indistinguishability matters mostly in long-haul cases (i.e. whether a longer/shorter fiber is used). Even inside 50km locality RTT depends on such factors as number of intermediary devices, which indirectly depends on the distance. Also, further improvement of technology may lower zero RTT so i18y distance will collapse further.

Also, if we'll move to the case of routing among wireless devices (routing in pervasive, mesh networks) then larger distance mean more energy consumption thus the distance factor becomes much more important.

So distance still matters 
and the geometrically shortest path is the preferred one in most cases. (At least for a simple model.)

\begin{figure}[h]
\center
\begin{tabular}{|c|c|r|}
\hline
source	&	destination	&	~approx.~RTT~ \\
\hline
UK	&	Australia	&	315-325 ms	\\
UK	&	Hong-Kong	&	300-335 ms	\\
~Yekaterinburg ~& Australia &	370 ms	\\
Hong-Kong& Australia	&	180 ms	\\
Netherlands & Vienna &	25-32 ms \\
localhost&	same city	&	$<$5 ms \\
localhost&~	same LAN segment ~&	0.2 ms	\\
localhost&	localhost	&	0.1 ms	\\
\hline
\end{tabular}
\caption {Typical traceroute RTT times.} \label{tab:rtt}
\end{figure}

\subsection {Solidity}

Different from younger Internet times, today the planet is rather densely covered with Internet infrastructure. One may check European or North-American fiber infrastructure maps or any major city fiber optics route maps\cite{cg}.

A traceroute from LIPEX (London Internet Providers Exchange) to BCIX (Berlin Commercial Internet Exchange) takes 7 hops (see Fig.~\ref{lipex2bcix}). So, an average hop is $\frac {930km}{7} \approx 133$ kilometers. 
This example demonstrates that even transit routes passing through populated areas are likely to be interrupted with IP (level 3) devices at a step having an order of $d_{i18y}$. As far as I see it, there is a tradeoff between network flexibility and overhead of inserting level 3 devices. If a ``cable'' is interrupted with level 3 devices in steps having order of $d_{i18y}$ or larger, then the delay caused by devices ($\tau_{0}$) is several times smaller than the delay caused by distance itself ($\tau_{c}$).

So, current (transit) networks are rather dense in terms of populated territory coverage. (I don't address oceanic cables here.) Off course, non-transit infrastructure has much higher density.
\begin{figure}
\begin{tabular}{|rll|rrr|}
\hline
\multicolumn{6}{|c|} {traceroute to 193.178.185.9 (193.178.185.9), 30 hops max, 40 byte packets}\\
\hline
N & host name & ip & RTT1 & RTT2 & RTT3 \\
\hline
 1 & lipex1.bdr.rtr.caladan.net.uk &193.109.219.24&  1.310   &1.038  & 0.823 \\
 2 & 195.66.224.185 &195.66.224.185~&  9.658  & 8.974   &9.009 \\
 3 & p15-0.core01.a03.atlas.cogentco.com &130.117.1.226&  18.384 &  9.530  & 8.875 \\
 4 & p5-0.core01.dus01.atlas.cogentco.com &130.117.1.126&  12.969  & 13.103  & 13.083 \\
 5 & p5-0.core01.ham01.atlas.cogentco.com~ &130.117.1.178&  18.455  & 18.024  & 18.150 \\
 6 & p6-0.core01.sxf01.atlas.cogentco.com &130.117.1.182&  21.388  & 21.286  & 21.032 \\
 7 & muli.bcix.de &193.178.185.9&  21.961  & 22.444  & 22.030 \\
 \hline
 \end{tabular}
\caption{LIPEX $\to$ BCIX traceroute (http://www.lipex.net/tools/)} \label{lipex2bcix}
\end{figure}
Once again, mesh/pervasive/internet-of-things networks may demonstrate dramatically higher densities than today's Internet -- due to participation of myriads of small/wireless/sensor devices.

Thus finally, the Euclidean model is (limitedly) applicable to the Internet.

\section {Continuous geo-aggregation theorem }

\subsection {On aggregation}

Aggregation is essential for scalable routing algorithms. Modern Internet routing protocols, such as OSPF \cite{ospf} and BGP \cite{bgp} employ aggregation schemes. OSPF uses area prefix aggregation, BGP uses CIDR aggregation since version 4. Limitedness of these schemes may be illustrated by the fact that one can not automatically determine even a continent a device resides in having its IP address. (Although, I do not claim that the problem has its roots in the routing protocols mentioned.)

The only popular routing scheme known to me that uses no aggregation at all is Dijkstra's shortest paths. It has computational complexity of $o(n^{2})$ so it does not scale well and it is obviously inapplicable in the continuous case (because the number of participants is infinite).

In the Euclidean model I will use the most straightforward method of spatial aggregation. I assume that it is possible to implement it, although it might not be true under contemporary patchy and area-centric address assignment policies. I will shed more light on this thesis in Section~\ref{sec:conc}.

\subsection {Surface, point-routers and covers}

Imagine Euclidean space where every point is a router. At least, those routers are placed so dense that we may think so. The purpose of the following routing algorithm is to transport a packet by the shortest path of length $l$ or, at least, by a path having stretch not larger than $\sigma$ (of length $< \sigma l$).
Packet forwarding is done in terms of directions and diminishingly small steps. A ``routed path'' is a curve drawn by a packet while it travels from source to destination.

One may ask why don't we use Cartesian coordinates to e.g. route by a straight line (Internet Coordinate System approach). Suppose we do not have such an ability (because we just pretend that the surface is solid). Points-routers may just remember \emph{directions} to some number of other point-routers.

To simplify the task of routing point-routers may use aggregates.
\begin{definition}{Aggregate} is an area (a convex) having one \emph{representative vertex}. If a sender does not know the shortest path to some destination, it forwards the packet by the shortest path to the representative vertex of the smallest aggregate containing the destination point.
\end{definition}
I assume that a point-router may trivially detect that a given point belongs to a given aggregate. Again, this thesis is presented in Sec.~\ref{sec:conc} in more detail.
\begin{definition}{Aggregate cover} is a set of aggregates covering all the populated space. If any point belongs to not more than $k$ aggregates of a given cover we call this cover $k$-fat.
\end{definition}
\begin{definition}{Symmetric cover} is a cover whose aggregates are identical geometric shapes (i.e. square grid or hexagonal grid - which have non-overlapping areas (1-fat), or just a cover formed by overlapping balls of the same radius).
\end{definition}
Just for the sake of simplicity I will further assume that we are dealing with a $k$-fat symmetric cover of a plane (2D) whose aggregates are balls of radius r.
\begin{definition} {Multilevel self-similar cover} is a stack of covers where each next cover is a scaled version of the previous one by a factor of $s$ (i.e. each cover is made of balls of radius $r_0s^i$, where $i$ is denoted as ``order'' of the cover)
\end{definition}

Each point-router assembles a cover of the populated space using aggregates of different orders from some global pre-given multilevel cover. For every aggregate used a point-router remembers direction to some representative point. These directions form ``routing table'' of the point-router.
Each point-router has a purpose of
\begin{enumerate}
\item using fewer aggregates
\item having a guarantee of stretch not larger than $\sigma$ for any routed path
\end{enumerate}
It is assumed that every router-point may forward by the shortest path if the destination point is closer than $r_0$. Stretch factor is understood as a ratio of the routed path length to the shortest path length (i.e. to the Euclidean distance between source and destination).

\subsection {Angular size lemma}

\begin{lemma}{(Aggregate's angular size lemma)}
Any routed path will have stretch under $\sigma$ if every point-router uses aggregates of angular size $\alpha \le \arccos \frac{1}{\sigma}$ (angular size is relative to the point-router).
\end{lemma}

Indeed, every point-router advances the "packet" to the representative vertex of the smallest aggregate containing the destination point by a diminishingly small step of length $\delta$. The angle between the direction to the representative point of the aggregate and the direction to the destination point is bounded by the maximum angular size of aggregates used by the forwarding point-router. Thus, the actual advance (i.e. packet-to-destination distance decrement) is not smaller than $\delta \cos \alpha$.   Obliously, integrating for all the path from the source to the destination we'll get that $\sigma < \frac{1}{\cos\alpha}$, i.e. the stretch is limited by the maximum angular size $\alpha$ of aggregates used by every point-router.

So, the restriction of using aggregates of angular size less than $\arccos \frac{1}{\sigma}$ guarantees the stretch of less than $\sigma$.

\subsection {The geo-aggregation theorem}

\begin{theorem}{Geo-aggregation theorem}

The number of aggregates consumed by a given point-router to assemble its cover depends logarithmically on the size of the space covered. In other words, in the described Euclidean space routing tables are logarithmically small.
\end{theorem}

\begin{figure}[ht]
\includegraphics[scale=0.4]{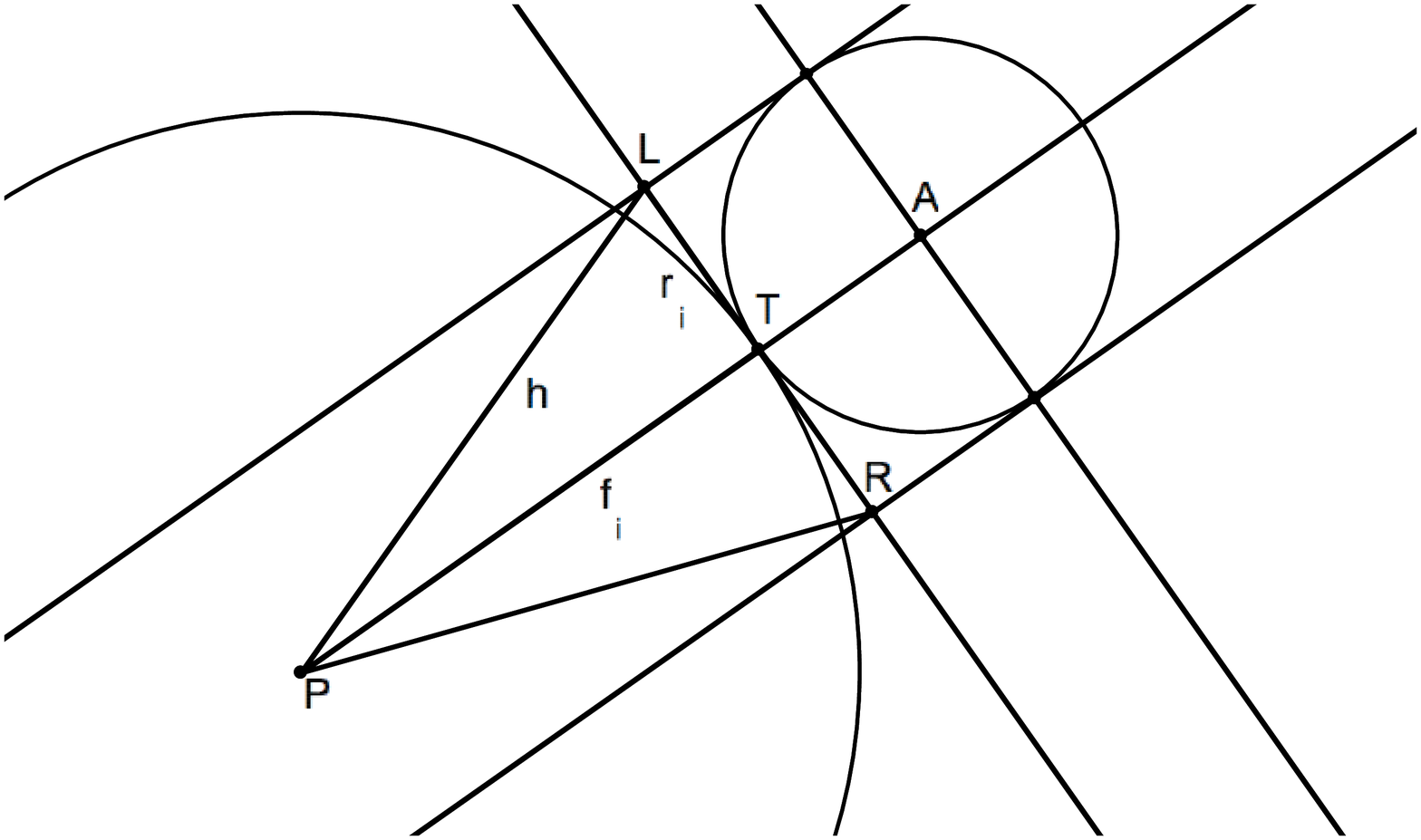}
\caption{How far an aggregate should be to fit into maximum angular size? $P$ is the point-router in question; the ball centered at $A$ is an aggregate of order $i$; lengths of $PT$ and $TR$ are equal to $r_{i}$; $\angle LPR$ is an upper estimate for the angular size of $A$.} \label{fig:faragg}
\end{figure}

\textbf{Proof.}
Router at point P may use ball aggregates of order $i$ if they are at least $f_i$ as far, where $f_{i} = \frac {r_{i}}{tg \alpha} = \frac {r_{0}} {tg \alpha} s^{i} $. Indeed, at that distance angular size of such a ball is under $\alpha$, see Fig.~\ref{fig:faragg}.

Starting from the distance of $f_{i}+2r_{i}$ the point-router may cover the space using aggregates of order not lower than $i$. Indeed, any point is covered by an aggregate of order $i$ and any such ball covering a point farther than $f_{i}+2r_{i}$ is itself farther than $f_{i}$, so $P$ may use it.

Thus, aggregates of orders lower than $i$ are needed only inside the ball centered at $P$ with a radius of $f_{i}+4r_{i}$. This ball has a square of 
\begin{equation}
B_{i} = \pi (f_{i}+4r_{i})^{2} =
\pi ( \frac {r_{0}} {tg \alpha} s^{i} + 4r_{0}s^{i})^2 =
\pi r_{0}^2 (\frac{1}{tg \alpha} + 4)^2 s^{2i}
\end{equation}
Thus, $P$ has to trace at most the following quantity of aggregates of order $i-1$:
\begin{equation}
n_{i-1} = \frac {B_i} {\pi r_{i-1}^2} k  = 
\frac { \pi r_0^2 ( tg^{-1} \alpha + 4)^2 s^{2i} } {\pi r_0^2 s^{2i-2}} k = 
(tg^{-1} \alpha + 4)^2 s^{2} k
\end{equation}
Thus, the number of aggregates taken from a given order $i$ is bounded by a constant and does not depend on $i$.

At the same time, the square of the space covered by aggregates of orders lower than $i$ grows as $B_{i}$, i.e. $\sim s^{2i}$.
Thus, to cover a World Ball of a given radius $R$ with a given stretch $\sigma$ we may use just $\sim \log_{s} R$ of orders in a multilevel self-similar cover and thus $\sim \log_{s} R$ of aggregates per point-router.  (Still the total number of aggregates used by all point-routers is not logarithmic.)

For nearly perfect routing having $\sigma=1+o$, where $o$ is small, $tg^{-1}\alpha \sim o^{-0.5}$. Thus the number of aggregates used by a router scales accordingly, $n_{i} \sim o^{-1}$. 

The problem of calculating optimal cover order scale factor $s$ for a given World Ball and $r_{0}$ is left as an exercise for the reader.
The interesting fact is that decreasing $r_{0}$ will cost us the same logarithmical price as increasing $R$.

\section {Conclusion}	\label{sec:conc}

Of course, the Euclidean model has limited applicability.
But, this model illustrates the fact that there is little technical sense for any Eurasian router to distinguish different destinations inside NYC the way BGP does.
Geo-aggregation with stretch of $1.1$ assumes that destinations as far as $10,000km$ might be grouped into aggregates of $4,000km$ each. 
This trivial assumption is in dramatic conflict with the current state-of-the-art in the Internet routing. It is my personal opinion that routing strategy overseeing continents is a bad strategy.

I will outline current obstacles that prevent any aggregation scheme having efficiency comparable to geo-aggregation. First, the current routing paradigm is area-centric by means that its top-level entities are autonomous systems which form some separate logical layer over IP addresses. Interrelationships of ASes/areas and IPs/routers/points have to be described and maintained in complex ways. Second, areas/ASes have more of organizational than of geographical underpinnings. Top-level routing (BGP) is AS-centric, so the most popular distance metric is seemingly AS hop which again has more of organizational than of time/space nature. Third, IP address assignments are patchy and again organization-centric.

Those problems are addressed not for the first time. IPv6 assignment policies are supposed to resolve IP address space fragmentation problem (``patchiness'') to some extent. RFC2374 \cite{rfc2374} specifies an approach of using IXes as IPv6 address assignment roots and thus exchange-based geographic aggregation.

It is my opinion that taking a point-centric approach may dramatically simplify the domain of Internet routing. As an example, the point-centric architecture uses no separate ``areas'' because each device just owns some prefixes and distributes sub-prefixes to downlink devices (to ``lesser aggregates'').
A~home page of the proposal is at http://www.topoip.org.


\begin{thebibliography}{99}
\bibitem{skitter} Bradley Huffaker, Marina Fomenkov, David Moore, and kc claffy: Macroscopic analyses of the infrastructure: measurement and visualization of Internet connectivity and performance, CAIDA
\bibitem{cg} http://www.cybergeography.org/ (Warning: most maps are 5 years old)
\bibitem{ospf} RFC 2328 -- OSPF Version 2
\bibitem{bgp} RFC 1771 -- A Border Gateway Protocol 4 (BGP-4)
\bibitem{rfc2374} RFC 2374 - An IPv6 Aggregatable Global Unicast Address Format
\end{thebibliography}
\end{document}